\documentclass[preprint2]{aastex}

\shorttitle{Supergranules as Probes of Solar Convection Zone Dynamics}
\shortauthors{Hathaway}

\begin{document}


\title{Supergranules as Probes of Solar Convection Zone Dynamics}
\author{David H. Hathaway}
\affil{NASA Marshall Space Flight Center, Huntsville, AL 35812 USA}
\email{david.hathaway@nasa.gov}

\begin{abstract}
Supergranules are convection cells seen at the Sun's surface as a space filling pattern of horizontal flows. While typical supergranules have diameters of about 35 Mm, they exhibit a broad spectrum of sizes from $\sim 10$ Mm to $\sim 100$ Mm. Here we show that supergranules of different sizes can be used to probe the rotation rate in the Sun's outer convection zone. We find that the equatorial rotation rate as a function of depth as measured by global helioseismology matches the equatorial rotation as a function of wavelength for the supergranules. This suggests that supergranules are advected by flows at depths equal to their wavelengths and thus can be used to probe flows at those depths. The supergranule rotation profiles show that the surface shear layer, through which the rotation rate increases inward, extends to depths of $\sim 50$ Mm and to latitudes of at least $70\degr$. Typical supergranules are well observed at high latitudes and have a range of sizes that extend to greater depths than those typically available for measuring subsurface flows with local helioseismology.
These characteristics indicate that probing the solar convection zone dynamics with supergranules can complement the results of helioseismology.
\end{abstract}


\section{INTRODUCTION}

Supergranules were discovered in the 1950s by \cite{Hart54} but were best characterized in the 1960s by \cite{Leighton_etal62} who gave them their name.
\cite{Leighton_etal62} showed that this cellular pattern of horizontal flows covers the solar surface and that the boundaries of the cells coincide with the chromospheric/magnetic network.
Typical supergranules have diameters of $\sim 35$ Mm and maximum flow speeds of $\sim 500$ m s$^{-1}$.
While the kinetic energy spectrum has a distinct peak at wavelengths of $\sim 35$ Mm, the spectrum includes cells at least three times larger and extends to much smaller cells where the supergranule spectrum blends into the granulation spectrum \citep{Hathaway_etal00}.

Larger cells live longer than smaller cells.
Typical supergranules with diameters of $\sim 30$ Mm live for $\sim 24$ hr \citep{SimonLeighton64, WangZirin89}.
Typical granules with diameters of $\sim 1$ Mm only live for $\sim 5$ min \citep{Title_etal89}.
Cells of intermediate size ($\sim 5-10$ Mm) have intermediate lifetimes of $\sim 2$ hr \citep{November_etal81}.

The rotation rate of the supergranule pattern was first measured by \cite{Duvall80} who cross-correlated the Doppler velocity pattern from equatorial spectral scans obtained over several days.
He found that the pattern rotates about 3\% faster than the photospheric plasma and faster rates are found for the 24-hr time lags from day-to-day than for the 8-hr time lags from the beginning to end of an observing day.
He concluded that larger cells dominate the longer time lags and that the observations are consistent with supergranules embedded in a surface shear layer in which the rotation rate increases with depth.

The presence of this shear layer was first suggested by \cite{FoukalJokipii75} as a consequence of the conservation of angular momentum by convective elements moving inward and outward in the near surface layers.
Global helioseismology inversions \citep{Thompson_etal96, Schou_etal98} indicate that the shear layer extends to depths of 35-50 Mm near the equator but may disappear or reverse at latitudes above $\sim 55\degr$.
While local helioseismology \citep{Basu_etal99, CorbardThompson02} does not probe as deeply, it produces similar results which suggest that the rotation rate may not continue to increase inward at higher latitudes.

\cite{BeckSchou00} measured the rotation of the supergranule pattern using a Fourier technique with space-based Doppler data from the ESA/NASA Solar and Heliospheric Observatory (SOHO) Michelson Doppler Imager (MDI) \citep{Scherrer_etal95}.
They mapped the data onto heliographic coordinates, took the Fourier transform in longitude of data from equatorial latitudes, and then the Fourier transform in time of those spectral coefficients over 6 10-day intervals in 1996 which had continuous coverage at a 15-min cadence.
They found that the larger cells do indeed rotate more rapidly than the smaller cells, but with rotation rates that exceeded the peak internal rotation rate at the base (50 Mm depth) of the surface shear layer as determined from global helioseismology \citep{Schou_etal98}.
This discovery led them to conclude that supergranules must have wave-like properties in order to rotate faster than the flows they are embedded in.

However, \cite{Hathaway_etal06} showed that line-of-sight projection effects on a rigidly rotating fixed velocity pattern could reproduce the excess rotation velocities found by \cite{BeckSchou00}.
In projecting the vector velocities onto the line-of-sight the function $\sin \phi$ (where $\phi$ is the heliographic longitude relative to the central meridian) multiplies the longitudinal flow velocities near the equator.
Since the flows are largely horizontal the $\sin \phi$ multiplier effectively pushes the peaks in the Doppler pattern away from the central meridian and makes the pattern appear to rotate more rapidly.

In fact, \cite{Schou03} largely removed the line-of-sight projection effects from the equatorial Doppler data by dividing the data by a function that approximated the function $\sin\phi$ and found that rotation velocities were much more in line with those from global helioseismology (but noted that there were still motions relative to this that suggested wave-like properties for supergranules).
Recently, \cite{Hathaway_etal10} found that the rotation profiles as functions of latitude determined by the cross-correlation technique used by \cite{Duvall80} for time lags from 2-hr to 16-hr could be reproduced by cellular patterns that are advected by a differential rotation with a peak velocity consistent with that found in the Sun's surface shear layer by global helioseismology.

Here we measure the rotation of the pattern of supergranules by analyzing the same data used by \cite{BeckSchou00} and by \cite{Schou03}.
We execute a series of 2D Fourier transform analyses.
We remove the line-of-sight projection effects near the equator as was done by \cite{Schou03} and repeat the Fourier analysis done by \cite{BeckSchou00} to show that the equatorial rotation rate of supergranules as a function of wavelength matches the equatorial rotation rate as a function of depth determined from global helioseismology \citep{Schou_etal98}.
This ``de-projection'' can only be done at the equator and is only approximate since the flows are not purely horizontal.
We determine the rotation rates at other latitudes by repeating the Fourier transform analysis on the raw Doppler data (without the removal of projections effects) and using a data simulation to support our conclusions.

\section{DATA PREPARATION}

The data consist of $1024^2$ pixel images of the line-of-sight velocity determined from the Doppler shift of a spectral line due to the trace element nickel in the solar atmosphere.
The images are acquired at a 1 min cadence and cover the full visible disk of the Sun.
We average the data over 31 min with a Gaussian weighting function which filters out variations on time scales less than about 16 min, and sample that data at 15 min intervals.
We then map these temporally filtered images onto a $1024^2$ grid in heliographic latitude from pole to pole and in longitude $\pm90\degr$ from the central meridian (Figure 1). This mapping accounts for the position angle of the Sun's rotation axis relative to the imaging CCD and the tilt angle of the Sun's rotation axis toward or away from the spacecraft. Both of these angles include modification in line with the most recent determinations of the orientation of the Sun's rotation axis \citep{BeckGiles05, HathawayRightmire10}. We analyze data obtained during a 60 day period of continuous coverage in 1996 from May 24 to July 22.

Line-of-sight projection effects influence the results so we also generate and analyze simulated data to assist in our determination of the actual rotation as a function of latitude and depth in the Sun's outer convection zone.
We construct the simulated data from an evolving spectrum of vector spherical harmonics in such a manner as to reproduce the spatial, spectral, and temporal behavior of the observed cellular flows.
The amplitudes of the spectral coefficients are constrained by matching the observed velocity spectrum \citep{Hathaway_etal00} with the radial flow component constrained by the disk center to limb variation in the RMS Doppler signal\citep{Hathaway_etal02}.

\begin{figure}[tbp]
\centerline{\includegraphics[width=\columnwidth]{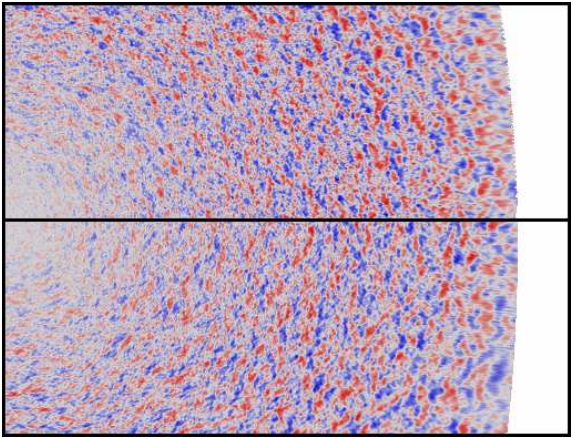}}
\caption{
Heliographic map details of the line-of-sight (Doppler) velocity from SOHO/MDI (top) and from the data simulation (bottom).
Each map detail extends $90\degr$ in longitude from the central meridian on the left and about $35\degr$ in latitude from the equator (the thick horizontal line).
The mottled pattern is the Doppler signal (blue for blue shifts and red for red shifts) due to the supergranules.}
\end{figure}

The cells are given finite lifetimes and made to rotate by adding changes to the phases of the spectral coefficients \citep{Hathaway_etal10}.
The rotation rates are constrained by matching the observed rotation rates as functions of latitude and wavelength.

The rotation rates of the cells in the simulated data are given by a fairly simple function with the latitudinal, $\theta$, variation separated from the wavelength, $\lambda$, variation such that
\begin{equation}
\Omega(\theta,\lambda)/2\pi = f(\theta) [1 + g(\lambda)]
\end{equation}
\noindent
with
\begin{equation}
f(\theta) = 454 - 51 \sin^2 \theta - 92 \sin^4 \theta\ \rm{nHz}
\end{equation}
\noindent
and
\begin{equation}
g(\lambda) = 0.045 \tanh {\lambda \over 31}\left[2.3 -\tanh{(\lambda - 65)\over 20}\right]/3.3
\end{equation}
\noindent
where the wavelength, $\lambda$, is given in Mm.

\section{EQUATORIAL ROTATION RATE}

We determine the equatorial rotation rate of the supergranules by 2D Fourier transforms
with and without removing line-of-sight projection effects.
The line-of-sight projection effects can be minimized using the method described by \cite{Schou03}.
The mapped Doppler velocities near the equator are divided by a function,
${\rm sgn}(\phi)\sqrt(\sin^2 \phi + 0.01),$
which approximates the geometric factor, $\sin \phi$, that multiplies the longitudinal velocity in producing the Doppler signal.
This ``de-projected'' signal and the raw Doppler signal are both then apodized near the limb and then Fourier transformed over longitude for the 50 latitude positions that straddle the equator.
These spectral coefficients are then Fourier transformed in time over six 10-day intervals.

The rotation rate as a function of wavelength is determined by first finding the temporal frequency of the centroid of the spectral power for that wavelength.
This temporal frequency is then divided by the longitudinal wavenumber to give the synodic rotation rate which is then converted to a sidereal rotation rate by adding a correction based on the rate of change of the ecliptic longitude during the observations ($\sim 1\degr \ {\rm day}^{-1}$).

The results of these analyses are shown in Fig. 2 along with the rotation rate with depth from a global helioseismology analysis by \cite{Schou_etal98}.
The sidereal rotation rate of the de-projected supergranules as a function of longitudinal wavelength matches the rotation rate as a function of depth through the outer half of the solar convection zone and the simulation matches the MDI observations.
The raw Doppler data give faster rotation rates than the de-projected data at all wavelengths (in both the MDI and the simulated data) with larger increases for larger cells.

\begin{figure}[tb]
\centerline{\includegraphics[width=0.7\columnwidth]{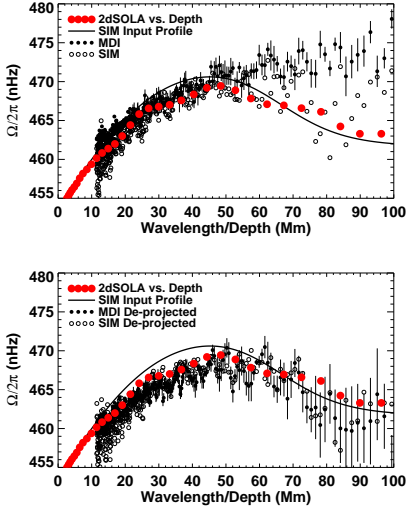}}
\caption{
The equatorial rotation rate for the Doppler pattern as a function of longitudinal wavelength.
Results for the raw data are shown in the upper panel.
Results for the de-projected data are shown in the lower panel.
MDI results are shown by the black dots (with $2 \sigma$ error bars for wavelengths longer than 20 Mm).
Simulated data results are shown by open circles.
The equatorial rotation rate as a function of depth from global helioseismology \citep{Schou_etal98} is shown by the large red dots while the rotation profile used in the simulation is shown by the solid black lines.}
\end{figure}

The rotation rates seen with both de-projected datasets fall slightly below both the helioseismology results {\em and} the input profile for the simulated data.
Experiments with the simulated data suggest that this can be attributed to the presence of a small radial flow component.
This component is projected into the line-of-sight along the equator by multiplying by a projection factor $\cos \phi$.
When the Doppler signal is de-projected by dividing by $\sin \phi$ this small signal can become large and influence the results in this manner.
 
We also see that the rotation rates for the raw simulated data fall systematically below the MDI results for wavelengths $> 50$ Mm.
This too may be a result of the radial flows.

These results do however lead to a key conclusion - that supergranules with sizes from 10 Mm to 100 Mm are advected by flows within the convection zone at depths equal to their widths.

\section{ROTATION PROFILES}

We determine the rotation rate of the supergranules as functions of latitude and wavelength (depth) by repeating the 2D Fourier transforms on the raw Doppler data for a series of latitude strips.
Each strip is 11 pixels or $\sim 2\degr$ high in latitude and offset from the previous strip by 5 pixels.
We repeat the procedure for 172 positions between $\pm 75\degr$ latitude and for each of six 10-day intervals for both MDI and simulated data.
Latitudinal rotation profiles are obtained for a series of cell wavelengths by averaging the profiles for all waveumbers that produce wavelengths within 5 Mm of the target wavelength.
These profiles are then averaged between hemispheres and smoothed with a 9-point binomial smoothing kernal which then limits the data to $\pm 70\degr$ latitude.
The results for averages from the MDI datasets and the from simulated data are shown in Fig. 3.

\begin{figure}[tb]
\centerline{\includegraphics[width=0.7\columnwidth]{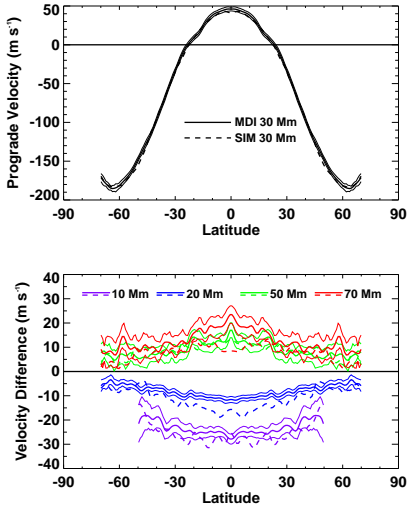}}
\caption{
The differential rotation profiles (average longitudinal velocity relative to a frame of reference rotating at the Carrington sidereal rate of 456 nHz) for the cells with 30 Mm wavelengths are shown in the upper panel with a solid line for the MDI data, a dashed line for the simulated data, and $2\sigma$ error limits on the MDI data indicated by thin solid lines.
The differences between the differential rotation at the other wavelengths from that found with the 30 Mm wavelength cells are shown in the lower panel using colors for the different wavelengths - violet for 10 Mm, blue for 20 Mm, green for 50 Mm, and red for 70 Mm.
}
\end{figure}

The rotation velocity {\em appears} to increase at all latitudes with increasing wavelength.
However, the small increase at 70 Mm wavelength (in particular near the equator) can be attributed to projection effects.
The actual rotation rate of the convection cells in the simulation is given by equations 1-3 and the solid lines in Fig 2. which give a rotation rate that increases with wavelength to a maximum at $\sim50$ Mm at all latitudes.

The smallest cells, $10\pm 5$ Mm, are difficult to resolve at high latitudes but clearly show slower rotation than larger cells over their observed latitude range.
Cells with wavelengths of $20 \pm 5$ Mm can be resolved at all latitudes and have significantly lower rotation rates than the larger cells.

While this method with the raw Doppler data is subject to systematic offsets due to line-of-sight projection effects, the same offsets are present in the simulated data and are much smaller for the smaller cells.
We conclude that the surface shear layer, in which the rotation rate increases inward, extends to latitudes of at least $70\degr$.

\section{CONCLUSIONS}

We conclude that supergranules are anchored or steered in the subsurface flows at depths equal to their wavelengths.
This is a simple explanation for the match with the rotation rate with depth from global helioiseismology (Figure 2 lower panel) and is based on well known physics.
This is consistent with numerical simulations of convection in the outermost 16 Mm of the Sun by \cite{Stein_etal11} who show that flow structures at different depths have diameters about equal to the depth itself.

This conclusion is, however, somewhat surprising given the much smaller estimates for the depth of typical supergranules previously determined from the visibility of their internal flows using local helioseismology.
\cite{Duvall98} estimated a depth of 8 Mm for typical supergranules while \cite{ZhaoKosovichev03} estimated a depth of 15 Mm.
This suggests that local helioseismology is less sensitive to these deeper (and slower) flows and that this new method of probing the convection zone with supergranules can probe flows at greater depths.

We also conclude that the surface shear layer extends to a depth of $\sim 50$ Mm at all latitudes.
The increase in rotation rate with depth has long been suggested by observations and is attributed to the conservation of angular momentum for fluid elements moving inward and outward in the near surface layers \citep{FoukalJokipii75, Hathaway82}.
Measurements of this rotation rate increase from helioseismology \citep{Schou_etal98, CorbardThompson02} indicate that it follows this critical gradient to depths of 10-15 Mm and reaches a maximum rotation rate at depths of 35-50 Mm. 
However, many helioseismology results also suggest that the shear layer disappears at latitudes above about $50\degr$.
The results reported here, using supergranules, indicate that the shear layer extends to the highest latitudes probed with this MDI data - $\sim 70\degr$.

The rotation increase with depth given by Equation 3 follows the critical gradient (with $\partial \ln \Omega / \partial \ln r = -1$) given by angular momentum mixing near the surface but then drops below that value at greater depths as the cell turn-over times become longer and the convective flows adjust to the solar rotation.

Global helioseismology can measure the internal rotation rate to great depths but it gives less reliable results at high latitudes.
Local helioseismology can measure the non-axisymmetric flows (as well as the axisymmetric meridional flow) but only in the near surface layers.
Using supergranules of different sizes to probe the flows in the Sun's convection zone extends these measurements to greater depths and higher latitudes.
This new method of probing solar convection zone dynamics should provide information complementary to that obtained with helioseismology.

\acknowledgements
The author would like to thank NASA for its support of this research through grants
from the Heliophysics Causes and Consequences of the Minimum of Solar Cycle 23/24 Program and the Living With a Star Program to NASA Marshall Space Flight Center. He is indebted to Ron Moore, Lisa (Rightmire) Upton, and an anonymous referee whose comments greatly improved the manuscript. He would also like to thank the American taxpayers who support scientific research in general and this research in particular. SOHO, is a project of international cooperation between ESA and NASA.

\end{document}